\documentclass{emulateapj}
\usepackage{apjfonts}

\begin{document}

\lefthead{IGR J11014--6103: A Hyper-Fast Pulsar?}
\righthead{Tomsick et al.}

\submitted{To appear in ApJ Letters}

\def\lsim{\mathrel{\lower .85ex\hbox{\rlap{$\sim$}\raise
.95ex\hbox{$<$} }}}
\def\gsim{\mathrel{\lower .80ex\hbox{\rlap{$\sim$}\raise
.90ex\hbox{$>$} }}}

\title{Is IGR J11014--6103 a Pulsar with the Highest Known Kick Velocity?}

\author{John A. Tomsick\altaffilmark{1},
Arash Bodaghee\altaffilmark{1},
Jerome Rodriguez\altaffilmark{2},
Sylvain Chaty\altaffilmark{2},
Fernando Camilo\altaffilmark{3},
Francesca Fornasini\altaffilmark{1}, \&
Farid Rahoui\altaffilmark{4,5}}

\altaffiltext{1}{Space Sciences Laboratory, 7 Gauss Way, 
University of California, Berkeley, CA 94720-7450, USA
(e-mail: jtomsick@ssl.berkeley.edu)}

\altaffiltext{2}{AIM - Astrophysique Interactions Multi-\'echelles
(UMR 7158 CEA/CNRS/Universit\'e Paris 7 Denis Diderot),
CEA Saclay, DSM/IRFU/Service d'Astrophysique, B\^at. 709,
L'Orme des Merisiers, FR-91 191 Gif-sur-Yvette Cedex, France}

\altaffiltext{3}{Columbia Astrophysics Laboratory, Columbia University, 
550 West 120th Street, New York, NY 10027-6601, USA}

\altaffiltext{4}{Harvard University, Astronomy Department, 60 Garden Street, 
Cambridge, MA 02138, USA}

\altaffiltext{5}{Harvard-Smithsonian Center for Astrophysics, 60 Garden Street,
Cambridge, MA 02138, USA}

\begin{abstract}

We report on {\em Chandra} X-ray and Parkes radio observations of 
IGR~J11014--6103, which is a possible pulsar wind nebula with a complex 
X-ray morphology and a likely radio counterpart.  With the superb 
angular resolution of {\em Chandra}, we find evidence that a portion 
of the extended emission may be related to a bow shock due to the 
putative pulsar moving through the interstellar medium.  The inferred 
direction of motion is consistent with IGR~J11014--6103 having been 
born in the event that produced the supernova remnant (SNR) MSH~11--61A.  
If this association is correct, then previous constraints on the 
expansion of MSH~11--61A imply a transverse velocity for 
IGR~J11014--6103 of 2,400--2,900\,km\,s$^{-1}$, depending on the SNR
model used.  This would surpass the kick velocities of any known 
pulsars and rival or surpass the velocities of any compact objects
that are associated with SNRs.  While it is important to confirm the 
nature of the source, our radio pulsation search did not yield a 
detection.

\end{abstract}


\section{Introduction}

IGR~J11014--6103 was discovered as a hard X-ray (20--100\,keV) source 
that was first seen during {\em INTEGRAL} \citep{winkler03} observations
of the Galactic Plane \citep{bird10}.  Approximately 500 ``IGR'' 
sources\footnote{See http://irfu.cea.fr/Sap/IGR-Sources.} have been 
discovered or detected for the first time in hard X-rays by {\em INTEGRAL}.
In the Galactic Plane, the {\em INTEGRAL} positions, which are typically 
accurate to a few arcminutes, are usually not sufficient to obtain source 
identifications.  Thus, the field was followed-up with {\em Swift} 
observations, and a soft X-ray counterpart was detected \citep{malizia11}.  

\cite{pavan11} report on {\em ROSAT}, {\em ASCA}, {\em Einstein}, {\em Swift}, 
and {\em XMM-Newton} observations of the region.  While IGR~J11014--6103 was
the target for {\em Swift}, the other satellites observed the region in order 
to study the supernova remnant MSH~11--61A (SNR~G290.1--00.8).
From the X-ray observations, \cite{pavan11} find that IGR~J11014--6103 has 
a complex X-ray morphology, consisting of a point source, an extended source 
that is $\sim$22$^{\prime\prime}$ North-East of the point source, and a streak 
of emission that extends to the North-West of the point source and is 
$\sim$4$^{\prime}$ in length. \cite{pavan11} also search optical/IR, radio, 
and gamma-ray catalogs for counterparts, and they report a possible 
association with the radio source MGPS-2 J110149--610104, which has a flux 
of $24.2\pm 4.8$\,mJy at 843\,MHz.  They suggest that IGR~J11014--6103 may 
be a pulsar wind nebula (PWN) produced by a high-velocity pulsar but 
conclude that X-ray observations with better angular resolution are 
required to establish the nature of the source.

Here, we report results of an observation with the {\em Chandra X-ray
Observatory}, which provides the improved angular resolution.

\section{{\em Chandra} Observation and Data Reduction}

A {\em Chandra} observation of IGR~J11014--6103 occurred on 2011 September 6 
(ObsID 12420).  We used the Advanced CCD Imaging Spectrometer 
\citep[ACIS,][]{garmire03} as the detector with the {\em INTEGRAL}
position of IGR~J11014--6103 at the ACIS-I aimpoint.  The observation
started at 1.43\,hr UT, and a net exposure time of 4,991\,s was obtained.

The data were initially processed at the {\em Chandra} X-ray Center
(CXC) with ASCDS Version 8.4.  After obtaining the data from the CXC,
we performed all subsequent processing with the {\em Chandra} 
Interactive Analysis of Observations (CIAO) version 4.3.1 software 
and Calibration Data Base (CALDB) version 4.4.6.  We used the CIAO 
program {\ttfamily chandra\_repro} to produce the ``level 2'' event
lists, which we analyzed as described below.

\section{Analysis and Results}

\begin{figure}
\includegraphics[clip,scale=0.45]{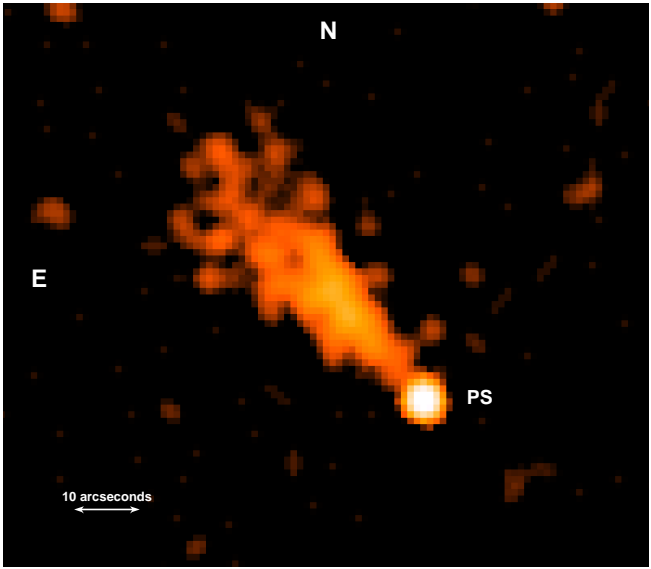}
\caption{{\em Chandra}/ACIS 0.3--10\,keV image of IGR~J11014--6103.  
There is a point source (PS) and an extension to the North-East.  
The image has $0^{\prime\prime}\!.98$ pixels and has been smoothed 
with a Gaussian kernel with a radius of $2^{\prime\prime}\!.95$.
\label{fig:j11014}}
\end{figure}

An inspection of the 0.3--10\,keV ACIS image indicates the presence of
the complex source that was previously reported, and the point source
(PS) and North-East (NE) extension are, by far, the brightest sources
in the field-of-view.  Figure~\ref{fig:j11014} shows NE in far more detail 
than in previous observations, indicating that the emission may, in fact, 
have a conical or cometary shape with its narrow end close to PS.  

\subsection{The Point Source}

We determined the {\em Chandra} position for PS using the CIAO routine
{\ttfamily wavdetect}, and it is R.A. = $11^{\rm h}01^{\rm m}44^{\rm s}.96$, 
Decl. = --$61^{\circ}01^{\prime}39^{\prime\prime}\!.6$ (equinox 2000.0)
with a 90\% confidence uncertainty of $0^{\prime\prime}\!.64$.  Although
the previously reported {\em XMM-Newton} position is marginally compatible 
with the $\sim$$0^{\prime\prime}.1$ (1-$\sigma$) near-IR position of 
2MASS J11014532--6101383, the {\em Chandra} position is 
$3^{\prime\prime}\!.0$ away, which rules out the association.  Furthermore, 
we searched all of the source catalogs available in the VizieR 
database\footnote{See http://webviz.u-strasbg.fr/viz-bin/VizieR-4.}
with a search radius of $2^{\prime\prime}\!.0$, and the {\em XMM-Newton} source
is the only one that appears.  Finally, we obtained near-IR images from
2MASS and optical images from the Digitized Sky Survey.  Although there 
is no evidence for near-IR or optical sources at the {\em Chandra} position
of PS, the nearby 2MASS source is relatively bright ($J = 11.2$, $I = 13.0$), 
making it difficult to obtain precise upper limits on the near-IR and optical
brightnesses without higher quality images.

\begin{figure}
\includegraphics[clip,scale=0.45]{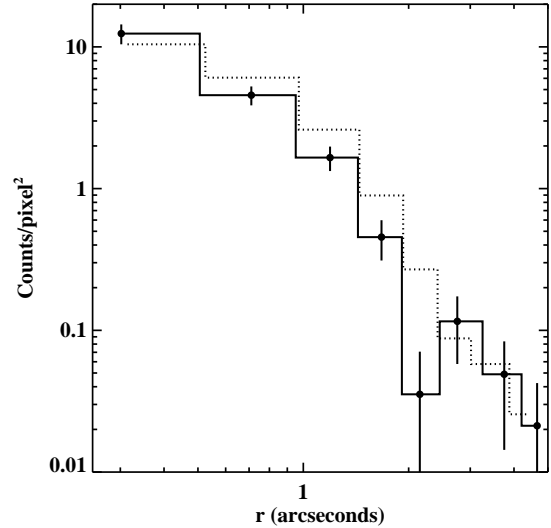}
\caption{The radial profile of PS ({\em solid line}).  The {\em dotted line} 
is the {\em Chandra} point spread function from ``ChaRT'' at the position 
of PS and with an energy spectrum like that of PS.\label{fig:rprofile}}
\end{figure}

We also produced a radial profile for PS to determine if it is, in fact,
a point source, and the profile is shown in Figure~\ref{fig:rprofile}.  
We used {\em Chandra} Ray-Tracing software (ChaRT) and the {\em Chandra}
simulator (``MARX'') to calculate the point spread function (PSF) at 
the position of PS.  The inputs to ChaRT were the off-axis angle
($3^{\prime}\!.64$), the azimuth ($330.82^{\circ}$), and the best fit spectrum
for PS described below.  Figure~\ref{fig:rprofile} shows that PS is
consistent with being a point source.  In fact, the simulated profile 
is slightly higher than the measurements for PS, and this is likely due to
inaccuracies within ChaRT or MARX as has also been found by \cite{bogdanov11}.

We extracted an energy spectrum for PS using the CIAO program 
{\ttfamily specextract}.  We used an aperture with a radius of 
$5^{\prime\prime}$ for the source spectrum, and a nearby rectangular 
source-free region for the background spectrum.  We extracted 126
counts in the source region in the 0.3--10\,keV band, and the 
background estimate (scaled to the size of the source region) is 
0.8 counts.  We fitted the energy spectrum using the XSPEC software 
package.  A model consisting of a power-law with interstellar 
absorption provides a good description of the spectrum.  The 
power-law photon index is $\Gamma = 1.5^{+0.6}_{-0.5}$ (90\% confidence 
uncertainties are given here and below), and the unabsorbed 0.3--10\,keV 
flux is $(8.6^{+3.6}_{-1.7})\times 10^{-13}$\,erg\,cm$^{-2}$~s$^{-1}$.  
Using \cite{wam00} abundances and \cite{bm92} cross sections, we 
obtained a column density of 
$N_{\rm H} = (1.6^{+0.9}_{-1.1})\times 10^{22}$\,cm$^{-2}$.  This is
consistent with the previous value measured with {\em XMM-Newton} 
\citep{pavan11}. 

We also checked for long and short-term variability of PS. Using the 
spectral shape described above, the {\em Chandra} observation indicates 
an absorbed 2--10\,keV flux of $5.5\times 10^{-13}$\,erg\,cm$^{-2}$~s$^{-1}$, 
and the flux obtained in the same band using the {\em XMM-Newton} 
observation is $(6.2^{+0.9}_{-2.6})\times 10^{-13}$\,erg\,cm$^{-2}$~s$^{-1}$ 
\citep{pavan11}.  Thus, there is no evidence for flux variations between 
the 2003 and 2011 observations.  Using {\em XMM-Newton}, \cite{pavan11} 
find no indication of variability on time scales of seconds to hours.  
We searched for variability with {\em Chandra} down to a time scale of 
50\,s using the CIAO tool {\ttfamily glvary} \citep{gl92} and did not
detect variations.

\subsection{The North-East Extension}

We extracted an energy spectrum for NE from a 
$24^{\prime\prime}$-by-$46^{\prime\prime}$ rectangular region rotated to
a position angle of $45^{\circ}$.  We chose the region to include 
all of the counts from NE and to exclude the counts from PS.  We 
extracted 261 counts in the source region in the 0.3--10\,keV band, 
and the background estimate (scaled to the size of the source region) 
is 11.7 counts.  A model consisting of a power-law with interstellar 
absorption provides a good description of the spectrum.  The 
power-law photon index is $\Gamma = 1.8\pm 0.4$ and the column 
density is $N_{\rm H} = (0.8^{+0.4}_{-0.3})\times 10^{22}$\,cm$^{-2}$.
The unabsorbed 0.3--10\,keV flux is 
$(1.3^{+0.4}_{-0.2})\times 10^{-12}$\,erg\,cm$^{-2}$~s$^{-1}$.  The spectral 
parameters are consistent with those found by \cite{pavan11}, but the 
flux measured by {\em Chandra} (the absorbed 2--10\,keV flux is 
$6.4\times 10^{-13}$\,erg\,cm$^{-2}$~s$^{-1}$) is a factor of $\sim$1.7 
higher.  It is possible that this difference reflects a true change 
in flux, but it may also be related to the fact that PS and NE are 
not fully resolved by {\em XMM-Newton}.

\begin{figure}
\includegraphics[clip,scale=0.48]{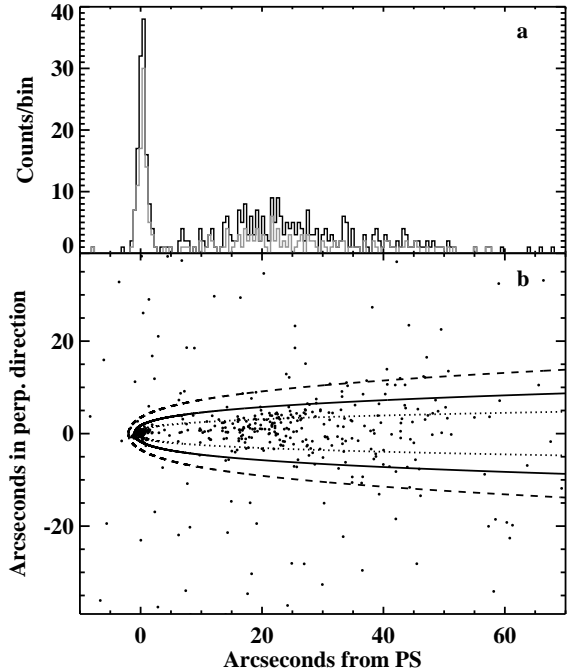}
\caption{{\em (a)} Counts profiles for PS and NE in two energy bands: 
the full 0.3--10\,keV band (black) and the 2--10\,keV band (grey).
The horizontal axis runs through PS and NE at a position angle of
$45^{\circ}$, and the counts in the perpendicular direction within 
$\pm$12$^{\prime\prime}$ of the axis are summed. {\em (b)} Unbinned 
0.3--10\,keV image where each dot corresponds to the location of an 
X-ray detection.  The dotted, solid, and dashed lines represent
the predicted locations of the bow shock for $\delta_{\rm SO} = 
0^{\prime\prime}\!.4$, $1^{\prime\prime}$, and $2^{\prime\prime}$, 
respectively, assuming $i = 0^{\circ}$.\label{fig:profile}}
\end{figure}

Profiles of detected counts for PS and NE are shown in 
Figure~\ref{fig:profile}a.  We extracted counts from a 
$24^{\prime\prime}$-by-$80^{\prime\prime}$ rectangular region rotated at a 
position angle of $45^{\circ}$ that includes both PS and NE.
We produced profiles of the counts along the long axis of the 
rectangle in two energy bands, 0.3--10\,keV and 2--10\,keV.  
While the number of 2--10\,keV counts is close to half the 
number in the full energy band, the profiles are not significantly
different.  The profile for NE peaks between 15$^{\prime\prime}$
and 30$^{\prime\prime}$ from PS.

\subsection{The X-ray Tail}

\begin{figure*}
\begin{center}
\includegraphics[clip,scale=0.3]{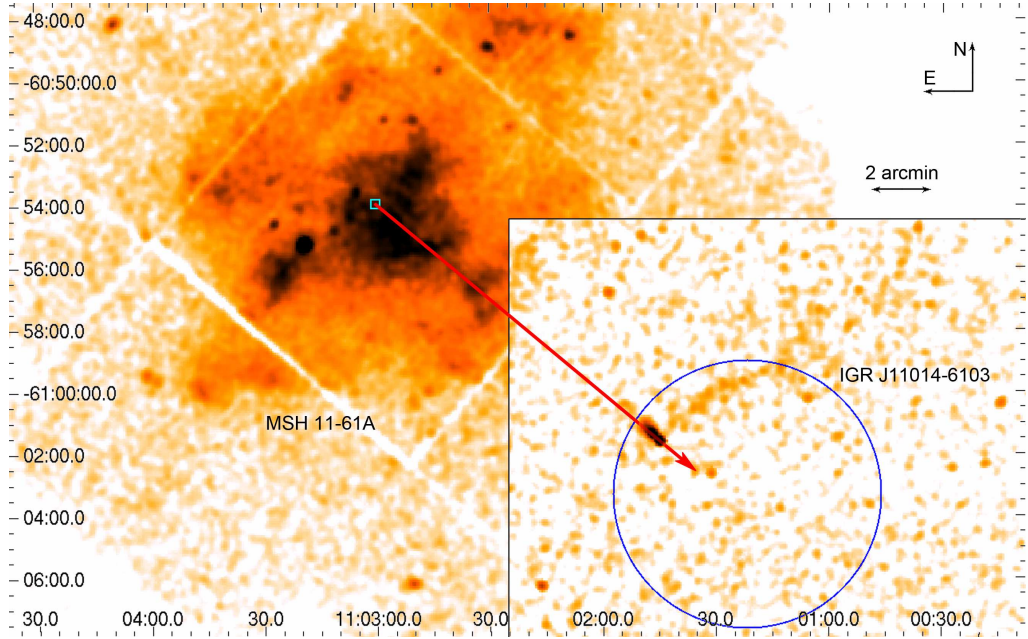}
\vspace{0.3cm}
\caption{Composite image of the supernova remnant MSH~11--61A near 
IGR~J11014--6103. The large image is from {\em XMM-Newton} (0.5--10\,keV), 
while the inset is from {\em Chandra}-ACIS (0.3--10\,keV): both were rebinned 
and smoothed. The blue circle is the $4^{\prime}.3$ {\em INTEGRAL} error 
circle.  It is shown via the red arrow that sources PS, NE, and the center 
of the SNR, marked with a cyan square, are co-aligned.  The 4$^{\prime}$ 
X-ray tail to the North-West of PS can also be seen.\label{fig:msh}}
\end{center}
\end{figure*}

In addition to PS and NE, the X-ray ``tail'' that was previously
reported from the {\em XMM-Newton} observations in \cite{pavan11} is 
also detected (see Figure~\ref{fig:msh}).  The length of the tail
is several arcminutes, and it is nearly perpendicular to the NE
extension. 

\section{Radio Pulsar Search}

We searched for pulsations from IGR~J11014--6103 on 2011 July 27 using
the CSIRO Parkes radio telescope.  We obtained a 5.6\,hr observation
with the 1.4\,GHz multibeam receiver and sampled every 1\,ms each of the 
96 frequency channels spanning 288\,MHz of bandwidth 
\citep[for further data acquisition details, see][]{manchester01}.  We 
analyzed the data following standard pulsar search techniques using PRESTO 
\citep{ransom01}, but did not identify a convincing pulsar candidate.  We
dedispersed the data up to $\mbox{DM} = 1280$\,pc\,cm$^{-3}$, which is 
twice the maximum predicted in this direction by the \cite{cl02} model
for the Galactic distribution of free electrons.  The sensitivity limit
of this observation, assuming a pulsar duty cycle of 10\%, is 
$S_{1.4} = 0.05$\,mJy, corresponding to a luminosity 
($L_{1.4} \equiv S_{1.4} d^2$) limit of $<$0.05\,mJy\,kpc$^2$ at a distance
of 1\,kpc and $<$5\,mJy\,kpc$^2$ at $d = 10$\,kpc.  If the source is at
the larger distance, the non-detection is not surprising since approximately 
half of all young pulsars detected in the radio have luminosities below 
5\,mJy\,kpc$^2$ \citep[see][]{camilo09}.

\section{Discussion}

The results presented above are consistent with (but do not prove) 
the hypothesis that PS is a pulsar.  The {\em Chandra} position 
shows that the source does not have a bright optical/IR counterpart, 
which is consistent with a pulsar since isolated neutron stars are 
very faint in the optical and IR.  The stability of the long-term 
flux and the power-law spectrum are also expected if the source is 
a pulsar.

The {\em Chandra} observation provides new information about the NE 
extension and its relationship to PS.  In the scenario where PS is a 
pulsar, the natural explanation for NE is that it is a PWN as suggested 
by \cite{pavan11}.  For some PWNe, a bow shock is created due to the 
motion of the pulsar through the ISM.  In the case of IGR~J11014--6103, 
the morphology of NE suggests that the pulsar may be moving in 
the South-West direction at high speed.  Figure~\ref{fig:msh} provides 
evidence in favor of this interpretation as it shows that NE is 
elongated in a direction that is co-aligned with PS and the center of the 
SNR MSH~11--61A.  Thus, a strong possibility is that the compact object 
that is powering IGR~J11014--6103 was produced when the supernova occurred.

MSH~11--61A has a centrally bright X-ray morphology, but the location of 
the compact object produced in the supernova is unknown.  The X-ray emission 
from the SNR is thermal rather than being powered by a PWN \citep{slane02}.  
It has been suggested that PSR~J1105--6107, which is $22^{\prime}$ from 
the center of the SNR, may be associated with MSH~11--61A \citep{kaspi97}.  
Applying SNR evolution models to MSH~11--61A, \cite{slane02} argue that 
the SNR is (1--2)$\times 10^{4}$\,yr old and is at a distance of 8--11\,kpc.  
The thermal conduction and cloudy ISM models would imply transverse
velocities for PSR~J1105--6107 of 4,500 and 5,300\,km\,s$^{-1}$, respectively 
\citep{slane02}.  In contrast, the known distribution of pulsar velocities 
includes groups near 175\,km\,s$^{-1}$ and 700\,km\,s$^{-1}$ \citep{cc98}.  
The fastest pulsar with parallax and proper motion measurements has a 
velocity of 1,100\,km\,s$^{-1}$ \citep{chatterjee05}.  Although there is
evidence for pulsar or compact object velocities in excess of 1,100\,km\,s$^{-1}$
\citep{hb06,wp07}, $>$1,000\,km\,s$^{-1}$ \citep{ng11}, and 
$>$1,400\,km\,s$^{-1}$ \citep{lovchinsky11}, the velocities implied by 
an association between PSR~J1105--6107 and MSH~11--61A are likely too high.

IGR~J11014--6103 is $11^{\prime}\!.9$ from the center of MSH~11--61A, and 
scaling from the calculations previously carried out for PSR~J1105--6107,
we determine transverse velocities of 2,400 and 2,900\,km\,s$^{-1}$ for the 
thermal conduction and cloudy ISM models, respectively, if IGR~J11014--6103
was formed in the supernova event that produced MSH~11--61A.  Among known 
compact objects associated with SNRs, these values are only rivaled by the 
recent estimate of 1,400--2,600\,km\,s$^{-1}$ for XMMU~J172054.5--372652
\citep{lovchinsky11}.  Thus, these two cases indicate that supernova models
that can produce $>$2,000\,km\,s$^{-1}$ kick velocities may be required.

\subsection{The PWN Bow Shock Interpretation and Implications}

If the interpretation that IGR~J11014--6103 is a PWN and that NE is 
emission from a bow shock is correct, then the fact that PS is a point 
source along with the morphology of NE provide a constraint on the angular 
distance between the pulsar and the standoff shock, 
$\delta_{\rm SO} = R_{\rm SO}\cos{i}/d$, where $R_{\rm SO}$ is the standoff
shock radius, $i$ is the inclination angle of the pulsar velocity 
with respect to the plane of the sky, and $d$ is the distance to the
source.  The standoff shock location is set by the ram pressure balance 
between the pulsar wind and the pulsar passing through the interstellar 
medium (ISM).  The shape of the entire bow shock in space is determined 
by $R_{\rm SO}$ according to 
\begin{equation}
R(\phi) = \frac{R_{\rm SO}}{\sin{\phi}} \sqrt{3(1-\frac{\phi}{\tan{\phi}})}~~~~,
\end{equation}
where $\phi$ is the angle with respect to the direction of motion, and
$R$ is the distance to the shock along each angle \citep{wilkin00}.  
While the observed shape of the bow shock will depend on $i$, 
here, we make a rough estimate for $\delta_{\rm SO}$ using the $i=0^{\circ}$ 
case.  Figure~\ref{fig:profile}b compares the calculated shape of the bow 
shock to the {\em Chandra} data for three values of $\delta_{\rm SO}$.  
The data are best described by a value close to $1^{\prime\prime}$, and the 
lower and upper limits are approximately $0^{\prime\prime}\!.4$ and 
$2^{\prime\prime}$, respectively.

Following \cite{caraveo03}, we derive 
\begin{equation}
\delta_{\rm SO} = 0^{\prime\prime}\!.266 \frac{\cos^{2}{i}}{v_{t,3}~d_{10}} (\frac{\dot{E_{36}}}{n_{0.1}})^{1/2}~~~~~,
\end{equation}
where $v_{t,3}$ is the transverse velocity of the pulsar in units of 
$10^{3}$\,km\,s$^{-1}$, $d_{10}$ is the distance to the pulsar in units
of 10\,kpc, $\dot{E_{36}}$ is the rotational energy loss of the pulsar
in units of $10^{36}$\,erg\,s$^{-1}$, and $n_{0.1}$ is the particle density
of the ISM near the source in units of 0.1\,cm$^{-3}$.

From the spectral fits described above, the 0.3--10\,keV unabsorbed flux 
of PS and NE combined is $2.2\times 10^{-12}$\,erg\,cm$^{-2}$\,s$^{-1}$, which 
corresponds to a luminosity of $2.6\times 10^{34}$ $d_{10}^{2}$\,erg\,s$^{-1}$, 
and as $\eta$ is the radiative efficiency of the PWN,
$\dot{E_{36}} = 0.026 d_{10}^{2} \eta^{-1}$.  We obtain another constraint 
by considering that if NE is a PWN, then the emission has a synchrotron 
origin and will decay as the electrons age on a time scale of $t_{s,3}$, 
where this is the exponential e-folding time for the X-ray flux in units 
of $10^{3}$\,yr.  Thus, the $\sim$$1^{\prime}$ transverse length of NE 
implies that $v_{t,3} = 2.8 d_{10} t_{s,3}^{-1}$, and the expression for the 
standoff shock radius becomes
\begin{equation}
\delta_{\rm SO} = 0^{\prime\prime}\!.015 \frac{\cos^{2}{i}~t_{s,3}}{d_{10}~n_{0.1}^{1/2}~\eta^{1/2}}~~~.
\end{equation}
A significant caveat is that the synchrotron lifetime constraint assumes 
that any velocity imparted to electrons in the PWN by the pulsar is small 
relative to the pulsar space velocity.  If NE has a magnetotail 
component \citep{rcl05}, then the ejection velocities could be high enough 
to invalidate this assumption \citep{kargaltsev08}.

\begin{figure}
\includegraphics[clip,scale=0.45]{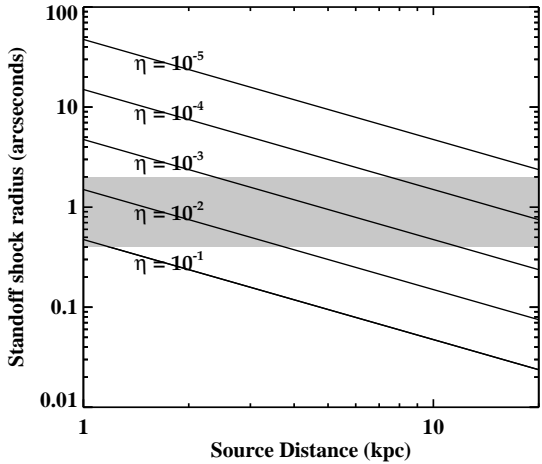}
\caption{Parameter space for the distance to IGR~J11014--6103 and the
standoff shock radius ($\delta_{\rm SO}$), if NE is indeed a bow shock.  
From the {\em Chandra} image, we find that values of $\delta_{\rm SO}$
between $0^{\prime\prime}\!.4$ and $2^{\prime\prime}\!.0$ are allowed
(for $i = 0^{\circ}$).  The solid lines show the calculated values for 
a range of pulsar efficiencies ($\eta$) for the case where $i = 0^{\circ}$, 
$n_{0.1} = 1$, and $t_{s,3} = 1$.\label{fig:pspace}}
\end{figure}

Figure~\ref{fig:pspace} shows the distance-$\delta_{\rm SO}$ parameter 
space for the case of $i = 0^{\circ}$, $t_{s,3} = 1$, and $n_{0.1} = 1$ 
for several different values of $\eta$ that span values measured for
$\sim$60 PWN \citep{kp10}.  The value of $t_{s,3} = 1$ is obtained in
the \cite{caraveo03} study, and it is what would be predicted for a
$\sim$10\,$\mu$G magnetic field \citep{kargaltsev08}.  The value of
$n_{0.1} = 1$ is consistent with the intercloud ISM density found in the 
vicinity of MSH~11--61A using the cloudy ISM model \citep{slane02}.  
The {\em Chandra} constraints on $\delta_{\rm SO}$ suggest a value of 
$\eta\sim 10^{-2}$ if the distance is $\sim$2\,kpc and values of 
$\eta\sim 10^{-3}$--$10^{-4}$ if the distance is $\sim$10\,kpc.  Although 
we cannot use this information to formally constrain the distance, the 
larger distance is slightly favored since PWNe with $\eta = 10^{-3}$--$10^{-5}$ 
are much more common than PWNe with larger radiative efficiencies \citep{kp10}.

\subsection{X-ray Tails from PWNe}

Whether IGR~J11014--6103 is related to MSH~11--61A or not, if PS is 
a pulsar moving in a South-West direction as implied by NE being a
bow shock, then the X-ray tail extends $\sim$$4^{\prime}$ in a direction 
that is $\sim$perpendicular to the motion of the pulsar.  A chance
alignment seems unlikely, especially since similar X-ray tails have been 
seen from other PWNe.  Examples include the Guitar Nebula (PSR~B2224+65)
and PSR~J0357+3205, which have $2^{\prime}$ and $9^{\prime}$ X-ray tails
\citep{hb07,deluca11}.  The Guitar Nebula is noteworthy because its 
X-ray tail points in a direction that is $118^{\circ}$ from the measured 
proper motion direction for this pulsar \citep{hb07,hui11}.  

While these examples support the PWN nature of IGR~J11014--6103 as
discussed by \cite{pavan11} and also the possibility that the 
IGR~J11014--6103 X-ray tail may extend in a direction perpendicular
to the pulsar's direction of motion, it is worth pointing out that
the size of the X-ray tail may be an argument against a $\sim$10\,kpc
distance.  While the X-ray tails for the other pulsars have lengths
near 1\,pc, a 10\,kpc distance for IGR~J11014--6103 would imply a 
$\sim$10~pc length for the X-ray tail from this source.

\section{Summary and Conclusions}

This work provides further evidence that IGR~J11014--6103 is a PWN.  
The {\em Chandra} position for PS eliminates the previously suggested 
2MASS counterpart, and the available near-IR and optical images do not 
indicate the presence of another counterpart, which is consistent with 
the faint levels of optical/IR emission expected from a pulsar.  Most 
significantly, the {\em Chandra} observation has provided new information 
about the morphology of the extended source to the North-East of PS 
(called NE in this work), and we argue that the most likely interpretation 
is that NE is shaped by a bow shock caused by PS moving in the South-West 
direction at a high rate of speed.  

Given the alignment of PS, NE, and the supernova remnant MSH~11--61A, 
we suggest the possibility that IGR~J11014--6103 is a pulsar that was
formed by the supernova event that produced MSH~11--61A.  If correct, 
this would imply that IGR~J11014--6103 is at the same distance as the 
SNR, 8--11~kpc, and also that the supernova imparted a kick velocity 
of at least 2,400--2,900\,km\,s$^{-1}$.  While we show that the {\em Chandra} 
constraint on the standoff shock makes a scenario with a large distance 
and a high velocity plausible, caution is required due to uncertainties 
about the nature of the source, its orientation in space ($i$), and its 
environment.  One possible counterargument to IGR~J11014--6103 having
a distance compatible with the SNR is that this would imply a 10\,pc-long 
X-ray tail.  

\acknowledgments

We acknowledge useful discussions with J.~Halpern, V.~Kaspi, and 
S.~Boggs.  Support for this work was provided by NASA through 
{\em Chandra} Award Number GO1-12046X issued by the {\em Chandra} 
X-ray Observatory Center, which is operated by the Smithsonian 
Astrophysical Observatory under NASA contract NAS8-03060.  
The Parkes Observatory is part of the Australia Telescope, which 
is funded by the Commonwealth of Australia for operation as a 
National Facility managed by CSIRO.


\begin{thebibliography}{}

\bibitem[\protect\astroncite{{Balucinska-Church} \& {McCammon}}{1992}]{bm92}
{Balucinska-Church}, M., \& {McCammon}, D.,  1992, ApJ, 400, 699

\bibitem[\protect\astroncite{{Bird} et~al.}{2010}]{bird10}
{Bird}, A.~J., et~al., 2010, ApJS, 186, 1

\bibitem[\protect\astroncite{{Bogdanov} et~al.}{2011}]{bogdanov11}
{Bogdanov}, S., {Archibald}, A.~M., {Hessels}, J.~W.~T., {Kaspi}, V.~M.,
  {Lorimer}, D., {McLaughlin}, M.~A., {Ransom}, S.~M., \& {Stairs}, I.~H.,
  2011, ApJ, 742, 97

\bibitem[\protect\astroncite{{Camilo} et~al.}{2009}]{camilo09}
{Camilo}, F., {Ng}, C.-Y., {Gaensler}, B.~M., {Ransom}, S.~M., {Chatterjee},
  S., {Reynolds}, J., \& {Sarkissian}, J.,  2009, ApJ, 703, L55

\bibitem[\protect\astroncite{{Caraveo} et~al.}{2003}]{caraveo03}
{Caraveo}, P.~A., {Bignami}, G.~F., {De Luca}, A., {Mereghetti}, S.,
  {Pellizzoni}, A., {Mignani}, R., {Tur}, A., \& {Becker}, W.,  2003, Science,
  301, 1345

\bibitem[\protect\astroncite{{Chatterjee} et~al.}{2005}]{chatterjee05}
{Chatterjee}, S., et~al., 2005, ApJ, 630, L61

\bibitem[\protect\astroncite{{Cordes} \& {Chernoff}}{1998}]{cc98}
{Cordes}, J.~M., \& {Chernoff}, D.~F.,  1998, ApJ, 505, 315

\bibitem[\protect\astroncite{{Cordes} \& {Lazio}}{2002}]{cl02}
{Cordes}, J.~M., \& {Lazio}, T.~J.~W.,  2002, arXiv:astro-ph/0207156

\bibitem[\protect\astroncite{{De Luca} et~al.}{2011}]{deluca11}
{De Luca}, A., et~al., 2011, ApJ, 733, 104

\bibitem[\protect\astroncite{{Garmire} et~al.}{2003}]{garmire03}
{Garmire}, G.~P., {Bautz}, M.~W., {Ford}, P.~G., {Nousek}, J.~A., \& {Ricker},
  G.~R.,  2003,
\newblock in X-Ray and Gamma-Ray Telescopes and Instruments for Astronomy.
  Edited by Joachim E. Truemper, Harvey D. Tananbaum. Proceedings of the SPIE,
  4851, 28

\bibitem[\protect\astroncite{{Gregory} \& {Loredo}}{1992}]{gl92}
{Gregory}, P.~C., \& {Loredo}, T.~J.,  1992, ApJ, 398, 146

\bibitem[\protect\astroncite{{Hui} \& {Becker}}{2006}]{hb06}
{Hui}, C.~Y., \& {Becker}, W.,  2006, A\&A, 457, L33

\bibitem[\protect\astroncite{{Hui} \& {Becker}}{2007}]{hb07}
{Hui}, C.~Y., \& {Becker}, W.,  2007, A\&A, 467, 1209

\bibitem[\protect\astroncite{{Hui} et~al.}{2011}]{hui11}
{Hui}, C.~Y., {Huang}, R.~H.~H., {Trepl}, L., {Tetzlaff}, N., {Takata}, J.,
  {Wu}, E.~M.~H., \& {Cheng}, K.~S.,  2011, arXiv:1112.5816

\bibitem[\protect\astroncite{{Kargaltsev} et~al.}{2008}]{kargaltsev08}
{Kargaltsev}, O., {Misanovic}, Z., {Pavlov}, G.~G., {Wong}, J.~A., \&
  {Garmire}, G.~P.,  2008, ApJ, 684, 542

\bibitem[\protect\astroncite{{Kargaltsev} \& {Pavlov}}{2010}]{kp10}
{Kargaltsev}, O., \& {Pavlov}, G.~G.,  2010, X-ray Astronomy 2009; Present
  Status, Multi-Wavelength Approach and Future Perspectives, 1248, 25

\bibitem[\protect\astroncite{{Kaspi} et~al.}{1997}]{kaspi97}
{Kaspi}, V.~M., {Bailes}, M., {Manchester}, R.~N., {Stappers}, B.~W., {Sandhu},
  J.~S., {Navarro}, J., \& {D'Amico}, N.,  1997, ApJ, 485, 820

\bibitem[\protect\astroncite{{Lovchinsky} et~al.}{2011}]{lovchinsky11}
{Lovchinsky}, I., {Slane}, P., {Gaensler}, B.~M., {Hughes}, J.~P., {Ng}, C.-Y.,
  {Lazendic}, J.~S., {Gelfand}, J.~D., \& {Brogan}, C.~L.,  2011, ApJ, 731, 70

\bibitem[\protect\astroncite{{Malizia} et~al.}{2011}]{malizia11}
{Malizia}, A., {Landi}, R., {Bassani}, L., {Bird}, A.~B.~A.~J., {Gehrels}, N.,
  \& {Kennea}, J.~A.,  2011, The Astronomer's Telegram, 3290

\bibitem[\protect\astroncite{{Manchester} et~al.}{2001}]{manchester01}
{Manchester}, R.~N., et~al., 2001, MNRAS, 328, 17

\bibitem[\protect\astroncite{{Ng} et~al.}{2011}]{ng11}
{Ng}, C.-Y., {Bucciantini}, N., {Gaensler}, B.~M., {Camilo}, F., {Chatterjee},
  S., \& {Bouchard}, A.,  2011, arXiv:1109.2233

\bibitem[\protect\astroncite{{Pavan} et~al.}{2011}]{pavan11}
{Pavan}, L., {Bozzo}, E., {P{\"u}hlhofer}, G., {Ferrigno}, C., {Balbo}, M., \&
  {Walter}, R.,  2011, A\&A, 533, A74

\bibitem[\protect\astroncite{{Ransom}}{2001}]{ransom01}
{Ransom}, S.~M.,  2001,
\newblock {\em Ph.D. thesis\/}, Harvard University

\bibitem[\protect\astroncite{{Romanova}, {Chulsky} \& {Lovelace}}{2005}]{rcl05}
{Romanova}, M.~M., {Chulsky}, G.~A., \& {Lovelace}, R.~V.~E.,  2005, ApJ, 630,
  1020

\bibitem[\protect\astroncite{{Slane} et~al.}{2002}]{slane02}
{Slane}, P., {Smith}, R.~K., {Hughes}, J.~P., \& {Petre}, R.,  2002, ApJ, 564,
  284

\bibitem[\protect\astroncite{{Wilkin}}{2000}]{wilkin00}
{Wilkin}, F.~P.,  2000, ApJ, 532, 400

\bibitem[\protect\astroncite{{Wilms}, {Allen} \& {McCray}}{2000}]{wam00}
{Wilms}, J., {Allen}, A., \& {McCray}, R.,  2000, ApJ, 542, 914

\bibitem[\protect\astroncite{{Winkler} et~al.}{2003}]{winkler03}
{Winkler}, C., et~al., 2003, A\&A, 411, L1

\bibitem[\protect\astroncite{{Winkler} \& {Petre}}{2007}]{wp07}
{Winkler}, P.~F., \& {Petre}, R.,  2007, ApJ, 670, 635

\end{thebibliography}

\end{document}